\documentclass[twocolumn,showpacs,preprintnumbers,amsmath,amssymb]{revtex4}

\usepackage{graphicx}% Include figure files
\usepackage{dcolumn}% Align table columns on decimal point
\usepackage{bm}% bold math

\begin{document}

\title{Shell-model interpretation of high-spin states in $^{134}$I}
% Force line breaks with \\

\author{L.~Coraggio$^1$}
\author{A.~Covello$^{1,2}$}
\author{A.~Gargano$^1$}
\author{N.~Itaco$^{1,2}$}

\affiliation{
$^1$Istituto Nazionale di Fisica Nucleare, Complesso Universitario di Monte
S. Angelo, I-80126 Napoli, Italy\\
$^2$Dipartimento di Scienze Fisiche, Universit\`{a} di Napoli Federico II, 
Complesso Universitario di Monte S. Angelo, I-80126 Napoli, Italy}

\date{\today}

\begin{abstract}
New experimental information has been recently obtained on the odd-odd nucleus 
$^{134}$I. We interpret the five observed excited states up to the energy 
of $\sim$3 MeV on the basis of a realistic shell-model calculation, and make
spin-parity assignments accordingly. A very good agreement is found between the
experimental and calculated energies.  
\end{abstract}

\maketitle

In a recent paper~\cite{Liu09}, excited levels up to an energy of about 3 MeV
were identified for the first time in $^{134}$I through the measurements of
prompt $\gamma$ rays from the spontaneous fission of $^{252}$Cf. A  five
transition cascade was observed, but the measured angular correlations
were not sufficient to assign spins and parities. In this connection, it may be 
mentioned that   preliminary results were also reported in Ref.~\cite{Mason09} 
on some new $\gamma$ transitions in $^{134}$I populated  in the reaction 
$^{136}$Xe + $^{208}$Pb. 
It is the aim of the present paper to give a shell-model interpretation of 
the observed levels.

The study of nuclei in the vicinity  of doubly magic $^{132}$Sn is indeed a
subject of  great current experimental and theoretical interest. Experimental
information on  these nuclei, which have been long inaccessible to
spectroscopic studies, is now becoming available offering the
opportunity to test shell-model  calculations in regions of shell closures 
off stability.

The odd-odd nucleus $^{134}$I with three protons and
one neutron hole away from $^{132}$Sn represents an important source of 
information on the matrix
elements of the proton-neutron hole interaction. Actually, the most 
appropriate system to study this interaction is $^{132}$Sb 
with only one proton valence particle and one neutron valence hole.
Experimental information on  this nucleus  was provided by the studies of
Refs.~\cite{Stone89,Mach95,Bhattacharyya01} and there have also been
various shell-model studies employing realistic effective 
interactions~\cite{Andreozzi99,Coraggio02,Brown05}.  Both the calculations  of
Refs.~\cite{Coraggio02} and \cite{Brown05} start from the CD-Bonn nucleon-nucleon ($NN$)
potential and derive the effective proton-neutron  interaction within the 
particle-hole formalism.  In the former paper, however, 
the short-range repulsion of the $NN$ potential is renormalized by means of
the $V_{\rm low-k}$ approach~\cite{Coraggio09} while in~\cite{Brown05} use is made of the traditional Brueckner $G$-matrix method.

The nucleus $^{134}$I, with an additional pair of protons with respect 
to $^{132}$Sb,  may
certainly  contribute to improve our knowledge of  the  two-body effective
interaction, since  it offers  the opportunity to investigate its  effects
when  moving away from the one proton-one neutron hole  system.
This  was also the motivation for   extending,  in Ref.~\cite{Coraggio02}, 
our shell model calculations to the nucleus $^{130}$Sb with  three neutron 
holes.

In our calculations we consider $^{132}$Sn as a closed core and let the valence
protons and neutron hole occupy the five levels $0g_{7/2}$, $1d_{5/2}$,
$1d_{3/2}$, $2s_{1/2}$, and $0h_{11/2}$ of the 50-82 shell. The single-particle
and single-hole energies have been taken from the experimental 
spectra~\cite{XUNDL} of
$^{133}$Sb  and $^{131}$Sn, respectively. The only exception is
the proton $\epsilon _{s_{1/2}}$ which has been  taken from 
Ref.~\cite{Andreozzi97}, 
since the corresponding single-particle level is still missing 
in the spectrum of $^{133}$Sb. Our
adopted values for the proton single-particle energies are (in
MeV):  $\epsilon_{g_{7/2}} = 0.0$, $\epsilon_{d_{5/2}} = 0.962$, 
$\epsilon_{d_{3/2}} = 2.439$, $\epsilon_{h_{11/2}} = 2.793$, and
$\epsilon _{s_{1/2}}= 2.800$, and for the neutron single-hole  energies:
$\epsilon^{-1}_{d_{3/2}} = 0.0,$  $\epsilon^{-1}_{h_{11/2}} = 0.065$,
$\epsilon^{-1}_{s_{1/2}} =0.332$, $\epsilon^{-1}_{d_{5/2}} = 1.655$, and 
$\epsilon^{-1}_{g_{7/2}} = 2.434$. Note that
for the $h_{11/2}$ level we have taken the position suggested in 
Ref.~\cite{Fogelberg04}.

As in our recent studies~\cite{Coraggio05,Coraggio06,Covello07,Gargano09} 
in the $^{132}$Sn region, we start from the CD-Bonn $NN$ 
potential~\cite{Machleidt01}
and derive the $V_{\rm low-k}$ with a value of the cutoff parameter 
$\Lambda=2.2$ fm$^{-1}$.
This low-momentum potential,  with the addition of the Coulomb force for
protons,  is then used to derive the effective interaction $V_ {\rm eff}$
within the framework of the $\hat Q$-box folded diagram 
expansion~\cite{Coraggio09} including diagrams up to second order
 in $V_{\rm low-k}$.
The computation of these diagrams is performed within the
harmonic-oscillator basis, using intermediate states composed of all
possible hole states and particle states restricted to the five shells
above the Fermi surface. This guarantees stability of the results when
increasing the number of intermediate states. The oscillator
parameter is $\hbar  \omega = 7.88$ MeV.

As mentioned above, the effective  proton-neutron interaction is derived
directly
in the particle-hole  representation, while for the  proton-proton interaction
we
use the
particle-particle formalism.  The shell-model calculations have been performed 
using the OXBASH computer code~\cite{Oxbash}.

\begin{figure}
\includegraphics[width=0.4\textwidth]{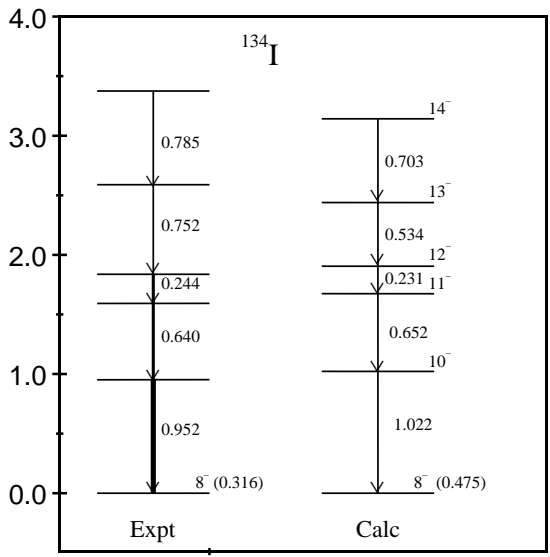}
\caption{\label{LevelScheme}
The five-transition cascade observed in $^{134}$I is compared with 
the calculated level scheme  for  negative-parity yrast states.}
\end{figure}

Let us now start to present our results  by comparing  in Table \ref{134Ilow} 
the calculated low-energy
spectrum  of $^{134}$I with  the experimental one \cite{ENSDF}. The levels 
shown in this table, which  have all been identified in the $^{134}$Te 
$\beta$ decay, were also observed in the experiment of Ref.~\cite{Liu09}. 
We see that the experimental energies are very well reproduced by theory,  the
largest discrepancy, 160 keV,  occurring for the $8^-$ state.  
For the two positive-parity states at 0.181 and 0.210 MeV
excitation energy our calculation speaks in favour of a $J=2$ and 3 
assignment, respectively.

As regards the wave functions, it turns out that the
six considered states are dominated by a single configuration, as is
shown  by the percentages reported in  Table~\ref{134Ilow}. It is 
interesting to note that these states may be viewed as the evolution of the 
six lowest-lying states in $^{132}$Sb, the $8^-$ state being tentatively
identified~\cite{ENSDF} with the 0.200 MeV level. As discussed in 
Ref.~\cite{Coraggio02}, the
first four positive-parity states in $^{132}$Sb are interpreted as 
members of the $\pi g_{7/2} \nu d^{-1}_{3/2}$ multiplet , while the
$3^+$ and  $8^-$  states as  the next to the highest $J$ member of the 
$\pi g_{7/2} \nu s^{-1}_{1/2}$and  $\pi g_{7/2} \nu h^{-1}_{11/2}$ multiplets,
respectively . The corresponding states in $^{134}$I are  dominated by the
same proton-neutron hole configuration with the remaining two
protons  forming a zero-coupled pair.  The main feature of  the 
proton-neutron hole multiplets, namely the lowest position of  the state 
with next to the
highest $J$, seems to be preserved  when adding two valence protons. We
find, however, that in $^{134}$I  the members of a given multiplet lie in a
smaller energy interval with respect to $^{132}$Sb, as is experimentally
confirmed  in the case of the  $\pi g_{7/2} \nu d^{-1}_{3/2}$ multiplet.

We now come to discuss the level scheme identified in Ref.~\cite{Liu09}, which
is reported in Fig.~\ref{LevelScheme} together with our shell-model 
interpretation.
The observed  $\gamma$ cascade is composed of five transitions and was
supposed to be built on the 0.316 MeV  $8^-$ isomeric state.  With this
assumption, we find that the excitation energies of the
experimental levels are well reproduced  by the calculated yrast sequence 
shown in the figure.
More quantitatively, the $\it {rms}$  deviation between the calculated 
and experimental energies is about 100 keV.

We have  verified that a different assumption for the lowest-lying populated 
level does not lead to any  theoretical sequence that matches well with  the
experimental energies and is consistent with the observed transitions. It
should be mentioned that  at about 0.250 MeV below the  $10^-$ state we
find the yrast $9^-$ state, whose probability to be  populated from the $10^-$
state is, however,  about 30 times smaller than that relative to  the 
$8^-$ state.
\begin{table}
\caption{\label{134Ilow}
Experimental and calculated energies (in keV) of the lowest lying states in 
$^{134}$I. Wave-function components with a percentage $\ge$ 10\%  are reported.}
\begin{ruledtabular}
\begin{tabular}{cccccc}
J$_{Exp}^{\pi}$ & E$_{Exp}$ & J$_{Th}^{\pi}$ & E$_{Th}$ & Configuration & Probability \\
\hline

(4)$^{+}$	&	0	&	4$^{+}$	&	0 & $\pi(g_{7/2})^{3}
\nu (d_{3/2})^{-1}$ & 78	\\
(5)$^{+}$	&     44	&	3$^{+}$     &	51 & $\pi(g_{7/2})^{3}
\nu (d_{3/2})^{-1}$& 77 	\\
(3)$^{+}$	&     79	&	5$^{+}$     &	122 & $\pi(g_{7/2})^{3}
\nu (d_{3/2})^{-1}$& 81	\\
(2,3)$^{+}$	&	181	&	2$^{+}$     &	158 & $\pi(g_{7/2})^{3}
\nu (d_{3/2})^{-1}$& 78	\\
(2,3)$^{+}$	&	210	&	3$^{+}$     &	312 & $\pi(g_{7/2})^{3}
\nu (s_{1/2})^{-1}$ & 60 	\\
& & & & $\pi g_{7/2})^{3} \nu (d_{3/2})^{-1}$ & 24 	\\
(8)$^{-}$	&     316	&	8$^{-}$     &	475& $\pi(g_{7/2})^{2}
 \nu (h_{11/2})^{-1}$ & 75	\\
& & & & $\pi g_7/2 (d_{3/2})^{2}  \nu (h_{11/2})^{-1}$ & 12 	\\
\end{tabular}
\end{ruledtabular}
\end{table}

In Table~\ref{134Ihigh}, we report the percentages of configurations larger 
than 10\% for the
$J^ \pi =$ $10^-$, $11^-$,  $12^-$, $13^-$, and $14^-$ states. As in the
case of the low-lying states (see Table~\ref{134Ilow}), each of
these high-spin states is  dominated by a single configuration.
We see that in all five states the neutron hole occupies  the $h_{11/2}$
level while  the three protons are in the $(g_{7/2})^3$   or $(g_{7/2})^2
d_{5/2}$ configurations, with two protons 
coupled to $J \ne 0$. In particular, the two highest-lying levels arise 
from the maximum spin alignment of the  corresponding configurations.

\begin{table}
\caption{\label{134Ihigh}
Wave functions  of negative-parity yrast states of $^{134}$I
(components $\ge$ 10 \% are reported).}
\begin{ruledtabular}
\begin{tabular}{ccc}
J$^{\pi}$ & Component & Probability  \\
\hline
    10$^{-}$ & $\pi (g_{7/2})^3 \nu h_{11/2}^{-1}$  & 89\\
    11$^{-}$ & $\pi (g_{7/2})^{2} d_{5/2} \nu h_{11/2}^{-1}$  & 92 \\
    12$^{-}$ & $\pi (g_{7/2})^{2} d_{5/2} \nu h_{11/2}^{-1}$  & 93 \\
    13$^{-}$ & $\pi (g_{7/2})^3 \nu h_{11/2}^{-1}$  & 66\\
    &  $\pi (g_{7/2})^{2} d_{5/2} \nu h_{11/2}^{-1}$  & 32 \\
    14$^{-}$ & $\pi (g_{7/2})^{2} d_{5/2} \nu h_{11/2}^{-1}$  & 97 \\
\end{tabular}
\end{ruledtabular}
\end{table}

In summary, we have given here  a shell-model description of $^{134}$I, 
focusing attention on the energy levels recently identified from the 
spontaneous fission of $^{252}$Cf~\cite{Liu09}. We have assigned spins and 
parity to the new 
observed levels and obtained a very good agreement between experimental and 
calculated energies. 
Our shell-model effective interaction has been derived from the CD-Bonn $NN$
potential  without using any adjustable parameter, in line with
our previous studies of other nuclei in the $^{132}$Sn region.
The accurate description obtained for all the investigated  nuclei makes us 
confident in the results of the present work.


\begin{thebibliography}{99}

\bibitem{Liu09} S. H. Liu {\it et al.},  Phys.Rev. C {\bf 79}, 067303 (2009).
\bibitem{Mason09} P. Mason {\it et al.},  Acta Phys. Pol. B {\bf 40}, 
489 (2009).
\bibitem{Stone89} C. A. Stone, S. H. Faller, and W. B. Walters, 
Phys. Rev. C {\bf 39}, 1963 (1989).
\bibitem{Mach95} H. Mach, D. Jerrestam, B. Fogelberg, M. Hellstr{\"o}m, J. P.
Omtvedt, K. I. Erokhina, and V. I. Isakov, Phys. Rev. C {\bf 51}, 500
(1995).
\bibitem{Bhattacharyya01} P. Bhattacharyya {\it et al.}, 
Phys. Rev. C {\bf 64},  054312 (2001).
\bibitem{Andreozzi99} F. Andreozzi, L. Coraggio, A. Covello, A. Gargano, 
T. T. S. Kuo,
and  A. Porrino, Phys. Rev. C {\bf 59}, 746 (1999).
\bibitem{Coraggio02} L. Coraggio, A. Covello, A. Gargano, N. Itaco, 
and T. T. S. Kuo, Phys. Rev. C {\bf 66}, 064311 (2002).
\bibitem{Brown05} B. A. Brown, N. J. Stone, J. R. Stone, I. S. Towner, and  
M. Hjorth-Jensen,  Phys. Rev. C {\bf 71}, 044317 (2005).
\bibitem{Coraggio09}
L. Coraggio, A. Covello, A. Gargano, T. T. S. Kuo and N. Itaco,
Prog. Part. Nucl. Phys. {\bf 62}, 135 (2009), and references therein.
\bibitem{XUNDL} Data extracted using the NNDC On-line Data Service from the 
XUNDL database, file revised as of 21 October 2009.
\bibitem{Andreozzi97}
F. Andreozzi, L. Coraggio, A. Covello, A. Gargano, T. T. S. Kuo and A. Porrino,
Phys. Rev. C {\bf 56}, R16 (1997).
\bibitem{Fogelberg04} B. Fogelberg {\it et al.} Phys. Rev. C {\bf 70}, 034312 
(2004).
\bibitem{Coraggio05} L. Coraggio, A. Covello, A. Gargano, and N. Itaco, Phys. 
Rev. C {\bf 72}, 057302 (2005).
\bibitem{Coraggio06} L. Coraggio, A. Covello, A. Gargano, and N. Itaco, Phys.
Rev. C  {\bf 73}, 031302(R) (2006).
\bibitem{Covello07} A. Covello, L. Coraggio, A. Gargano, and N. Itaco, Eur.
Phys. J. ST  {\bf 150}, 93 (2007).
\bibitem{Gargano09} A. Gargano, L. Coraggio, A. Covello, and N. Itaco, 
J. Phys.: Conf. Ser. {\bf 168}, 012013  (2009). 
\bibitem{Machleidt01} R. Machleidt, Phys. Rev. C {\bf 63}, 024001 (2001).
\bibitem{Oxbash} B. A. Brown, A. Etchegoyen, and W. D. M. Rae, the Computer
Code OXBASH, MSU-NSCL  Report No. 524.
\bibitem{ENSDF} Data extracted using the NNDC On-line Data Service from the 
ENSDF database, file revised as of 27 October 2009.


\end{thebibliography}
\end{document}